# Effects of Marine Protected Areas on Overfished Fishing Stocks with Multiple Stable States

by

Nao Takashina[1*] and Akihiko Mougi[2]

1. *Department of Biology, Faculty of Sciences,*
   *Kyushu University, 6-10-1, Hakozaki, Fukuoka, 812-8581, Japan*
   nao.takashina@gmail.com
2. *Department of Biological Sciences,*
   *Faculty of Life and Environmental Sciences,*
   *Shimane University, 1060 Nishikawatsu-cho, Matsue-shi, Shimane*
   *690-8504, Japan*
   amougi@gmail.com

**Abstract**

Marine protected areas (MPAs) have attracted much attention as a tool for sustainable fisheries management, restoring depleted fisheries stocks and maintaining ecosystems. However, even with total exclusion of fishing effort, depleted stocks sometimes show little or no recovery over a long time period. Here, using a mathematical model, we show that multiple stable states may hold the key to understanding the tendency for fisheries stocks to recover because of MPAs**.** We find that MPAs can have either a positive effect or almost no effect on the recovery of depleted fishing stocks, depending on the fish migration patterns and the fishing policies. MPAs also reinforce ecological resilience, particularly for migratory species. In contrast to previous reports, our results show that MPAs have small or sometimes negative effects on the recovery of sedentary species. Unsuitable MPA planning might result in low effectiveness or even deterioration of the existing condition.

**Keywords**: MPAs; depleted fishing stocks; regime shift; ecological resilience

---

[*] Corresponding author

# 1. Introduction

In recent years, the conditions of more than 30% of fishing stocks have been described as overexploited, depleted or recovering [1]. Marine protected areas (MPAs) have attracted much attention as a tool for sustainable fisheries management, restoring depleted fishing stocks and maintaining ecosystems [2-5]. A number of MPAs have been established around the world and have restored depleted fishing stocks [6-10]. However, depleted fishing stocks sometimes show little or no recovery over a long time period, despite the reduction or exclusion of fishing effort from the protected area [11]. This suggests that the creation of no-take marine reserves or MPAs might sometimes have no effect on the recovery of fishing stocks. One hypothesis to explain this lack of recovery is the existence of multiple stable states.

In natural environments, both terrestrial and marine, ecosystems often undergo catastrophic changes and the existence of multiple stable states has been asserted [12-15]. Historically, collapses of a number of fishing stocks due to the failure of fisheries management have been reported (e.g., [16]). When such damaged ecosystems recover, they often show the following characteristics of multiple stable states: (1) they exhibit varying degrees of hysteresis: the trajectories of recovery are different from that of decline [17]; (2) sustaining a resilient ecosystem is easier than recovering it after phase shift has occurred [17, 18]. In spite of recognizing the inherent patterns of multiple stable states in marine environments, many studies on MPAs have ignored the effects of multiple stable states. Consideration of multiple stable states provides us a new perspective to the effects of MPAs on depleted fishing stocks.

Here, using a mathematical model, we show the effects of MPAs on depleted fishing stocks having multiple stable states. In the following analysis, the term MPA is used referring to a no-take marine reserve where all human uses contributing ecosystem impacts are not allowed. First, we examine the effects of MPAs on the recovery of the equilibrium population size of the depleted fishing stock. Population size is the most common measure in fishery management, providing evidence for the efficacy of MPAs. Second, we examine the changes of ecological resilience, defined as the size of the basin of attraction or the width of the stability basin in a common ball-in-cup diagram [19]. This measure is not commonly used in fishery management, but it might provide a useful insight into how to manage a fishing stock with multiple stable states. We show that the introduction of the MPA affects the restoration of the depleted fishing stocks and can enhance the ecological resilience. We also show that the degree of these positive effects of MPAs varies widely depending on the migration characteristics of the target species and the fishing policies. The results provide us with new insights as to how the MPA works in the management of depleted stocks and in resilience-based ecosystem management [17, 20, 21].

## 2. Methods

*2.1. Population dynamics*

To explore the efficacy of MPAs in ecosystems with multiple stable states, we use the canonical model giving multiple stable states, which is often applied to marine systems (e.g., [12, 13, 22]):

$$\frac{dX}{dt} = rX\left(1 - \frac{X}{K}\right) - \frac{aPX^2}{b^2 + X^2}, \tag{1}$$

where $X$ is the population density, $r$ is the intrinsic growth rate, $K$ is the carrying capacity, $a$ is the predator's consumption rate, $P$ is the constant predator density, and $b$ is the half-saturation level of predation. By introducing a linear harvest term, which is commonly used in fisheries models, Eq. (1) becomes

$$\frac{dX}{dt} = rX\left(1 - \frac{X}{K}\right) - \frac{aPX^2}{b^2 + X^2} - qEX, \tag{2}$$

where $q$ is the catchability coefficient and $E$ is the fishing effort. $E$ is typically specified as the number of vessels actively fishing [23]. This model has been used to evaluate the effects of alternative stable states in fisheries and is known to have lower stable state $X_L^*$ and upper stable state $X_U^*$ [24, 25].

When the MPA is introduced, the ecosystem would be separated into two patches: the fishing ground and the protected area (MPA; Fig. 1). Hence, we use a two-patch model to explore the effects of MPAs (e.g., [25-31]). When the MPA is considered, the model is described by the following two equations (see Appendix A for more details):

$$\frac{dX_1}{dt} = rX_1\left(1 - \frac{X_1}{K}\right) - \frac{aPX_1^2}{b^2 + X_1^2} - \frac{qEX_1}{1 - \sigma R} + mR\left(\left(\frac{X_2}{K}\right)^s X_2 - \left(\frac{X_1}{K}\right)^s X_1\right), \tag{3a}$$

$$\frac{dX_2}{dt} = rX_2\left(1 - \frac{X_2}{K}\right) - \frac{aPX_2^2}{b^2 + X_2^2} + m(1-R)\left(\left(\frac{X_1}{K}\right)^s X_1 - \left(\frac{X_2}{K}\right)^s X_2\right), \tag{3b}$$

where $X_1$ and $X_2$ are the population density in the fishing ground and the MPA, respectively. $m$ is the migration rate. $R$ is the fraction of the MPA and $1-R$ is the fraction of the fishing ground. $\sigma$ is the effort redistribution coefficient which represents the intensity of fishing effort transfer into the fishing ground due to the creation of the MPA. When $\sigma = 0$, the fishing effort over the fishing ground does not change, with or without the MPA. In other words, the number of vessels actively fishing in the fishing ground per unit area is same as before the MPA creation (CEP: constant effort policy). When $0 < \sigma < 1$, the fishing effort increases with the increasing fraction of MPA. This corresponds to a situation where fishing vessels previously exerted in the pre-MPA are redistributed to the fishing ground and as a result the number of vessels actively fishing in the fishing ground per unit area is increased after the establishment of the MPA (ERP: effort redistribution policy), at the rate of inverse $1 - \sigma R$.

The last terms on the right hand side of Eqs. 3 represent migration between two patches involving the density effect defined by Amarasekare [32]. $s$ is the strength of density-dependence of migration. When $s = 0$, the migration is random. This case corresponds to the random migration defined, for example, by Takashina et al. [31]. The first and second terms in the square brackets of Eq. (3a) represent the immigration from the patch 2 and the emigration to the patch 2, respectively [32]. When $s > 0$, emigration increases with population density at an accelerating rate (density-dependent migration; DM). This pattern of migration has been documented in a number of marine organisms, including fish and echinoderms [33-35]. When $-1 < s < 0$ emigration increases with population density at a decelerating rate (negative density-dependent migration; NDM). This type of migration may occur as a result of the Allee effect [36].

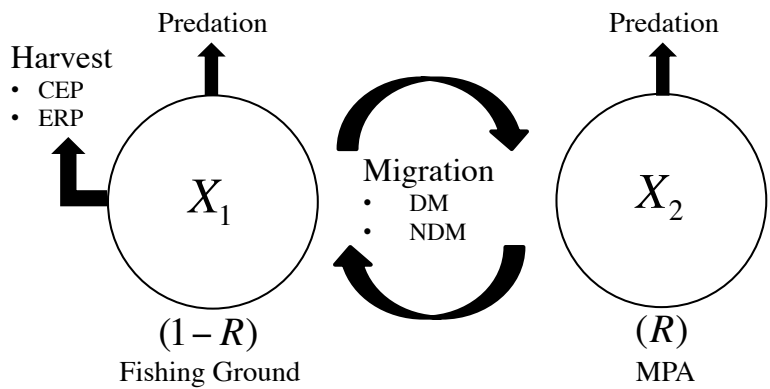

**Figure 1** Schematic description of the model.

To reduce the number of parameters, we use the following non-dimensional form of the equations [37] (See Appendix A):

$$\frac{d\hat{x}_1}{d\tau} = \hat{x}_1(1-\hat{x}_1) - \frac{\beta_2 \hat{x}_1^2}{\beta_1^2 + \hat{x}_1^2} - \frac{\beta_3 \hat{x}_1}{1-\sigma R} + \beta_4 R\left(\hat{x}_2^{s+1} - \hat{x}_1^{s+1}\right), \quad (4a)$$

$$\frac{d\hat{x}_2}{d\tau} = \hat{x}_2(1-\hat{x}_2) - \frac{\beta_2 \hat{x}_2^2}{\beta_1^2 + \hat{x}_2^2} + \beta_4 (1-R)\left(\hat{x}_1^{s+1} - \hat{x}_2^{s+1}\right), \quad (4b)$$

where, $\hat{x}_i$ ($i$ = 1, 2) is the non-dimensional parameter of $X_i$, and $\tau$ is the non-dimensional time metric scaled by intrinsic growth rate (see Appendix A). Here, we assume that multiple stable states are introduced due to the fishing activity. That is, multiple stable states do not exist when the MPA is absent ($R$=0). We set $\beta_1 = 0.1$ [24, 25], $\beta_2 = 0.17$, and $\beta_3 = 0.15$. These parameter values are chosen so as to have multiple stable states in our model, in the absence of MPA ($R$ = 0).

*2.2. Ecological resilience*

In the presence of multiple stable states, we can consider the ecological resilience, which is the 2-dimensional extension of the ecological resilience defined by Peterson *et al.* [19]: the length of basin of attraction. Here we define ecological resilience in a 2D plane as $\alpha_{R,80\%}/\alpha_R$, where $\alpha_R$ is an area of a phase plane of Eqs. 4 with a MPA of a fraction $R$. $\alpha_{R,80\%}$ is the area that any initial point in the region converges toward a stable equilibrium retaining $\geq 80\%$ of the population size of the uppermost stable equilibrium of Eqs. 4; $\hat{x}_1^U + \hat{x}_2^U$, where $\hat{x}_1^U$ and $\hat{x}_2^U$ are the uppermost stable equilibrium of Eqs. 4. For the natural extension of the 1- to 2-dimensional ecological resilience, the region of $\alpha_R$ and $\alpha_{R,80\%}$ are restricted by $0 < \hat{x}_1 < \hat{x}_1^U$ and $0 < \hat{x}_2 < \hat{x}_2^U$. In the absence of a MPA ($R$=0), our definition of the ecological resilience becomes $\alpha_{0,80\%}/\alpha \approx (\hat{x}^U - \hat{x}^M)/\hat{x}^U$, where $\hat{x}^U$ is the non-dimensional form of the upper stable equilibrium and $\hat{x}^M$ is the non-dimensional form of the equilibrium that separates two stable equilibriums, namely the separatrix of Eq. 2. This corresponds to the 1-demensional ecological resilience defined by Peterson [19]

**3. Results**

*3.1. Effects of the MPA on the equilibrium population size of depleted fishing stocks*

Consider a situation where the fishing stock is depleted and in a lower stable state because of overfishing. Here we introduce the MPA to restore the fishing stock. We set the initial conditions of the system as $\hat{x}_1^0 = (1-R) \times \hat{x}^L$, $\hat{x}_2^0 = R \times \hat{x}^L$, where $\hat{x}_1^0$ and $\hat{x}_2^0$ are the initial condition of $\hat{x}_1$ and $\hat{x}_2$, respectively. When CEP is applied, the introduction of the MPA can restore the equilibrium fishing stock drastically or gradually, or can have almost no effect (Fig. 2). The tendency of recovery due to the MPA depends on the migration rate (Fig. 2a). When the migration rate is either very small ($\beta_4 \leq 0.1$) or relatively large, a large MPA is necessary for sufficient recovery. When the migration rate is relatively small, the MPA has a large effect on the recovery. The MPA tends to be more effective when the density dependence of migration is strong ($s = 2.5$; Fig. 2b). The information for the plausible equilibrium is provided in Figs. A1 and A2 in Appendix.

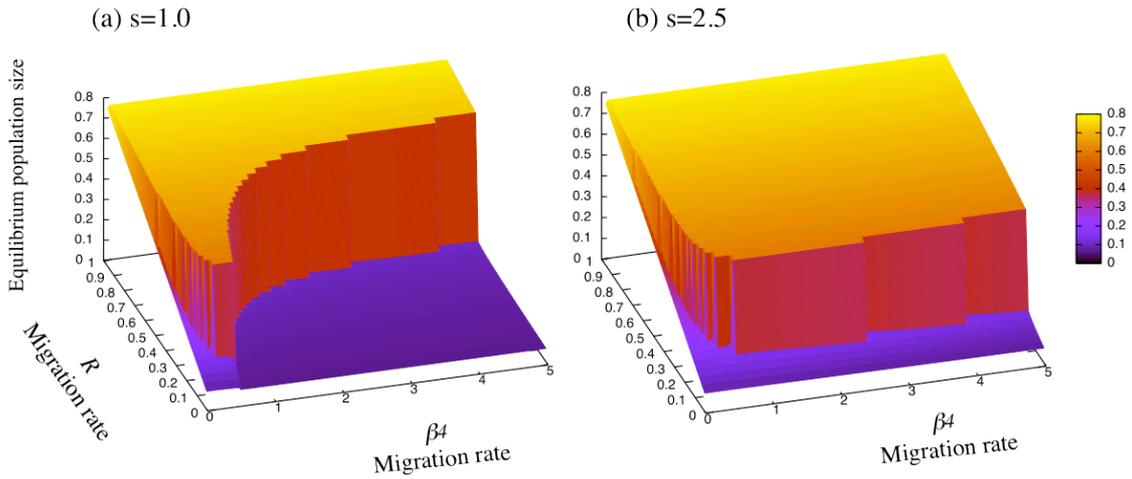

**Figure 2** Effects of introducing the MPA on equilibrium population size. Colors represent equilibrium population sizes. We assume CEP ($\sigma = 0$) and DM ($s > 0$). The strength of density-dependent migration differs in (a) and (b). The recovery of equilibrium population size due to the MPA depends on whether the dynamics starting from initial population size converge to upper stable equilibrium or not.

The effects of the MPA depend not only on the manner of migration but also on fishing policy. Here, for convenience, we define "small recovery" as an increase in equilibrium population size of less than 10% of the uppermost stable equilibrium population: $\leq \hat{x}^L + 0.1\hat{x}^U$, where $\hat{x}^L$ is the non-dimensional form of the lower stable equilibrium of Eq. 2. Figure 3 shows the relative size of parameter space showing small recovery. Here parameter space is restricted by $0 < R < 1.0$ and $0 < \beta_4 < 5.0$ as in Figs 2. ERP is less likely to recover the equilibrium population size compared with CEP as shown in Fig. 3. However, when migration manner is strongly density-dependent (larger values of *s*), the recovery is likely, regardless of fishing policies.

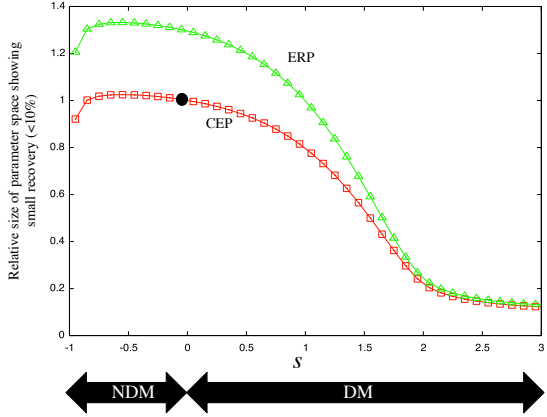

**Figure 3** Relative size of parameter space showing small recovery with varying migration manners, *s*. The black circle represents the reference point, $s = 0$ on the CEP curve.

*3. 2. Effects of the MPA on the ecological resilience*

The introduction of the MPA also greatly affects the ecological resilience. We consider the relative ecological resilience, $\left(A_{R80} / A_R\right) / \hat{A}$, to be 1 in the absence of the MPA.

Under CEP, the relative ecological resilience increases as the fraction of the MPA increases, regardless of the migration manners and rate (Fig. 4). However, under ERP, the MPA can have either a negative or positive effect on the ecological resilience depending on the migration rate. When the target species has a very low migration rate ($\beta_4 = 0.1$), ERP reduces the relative ecological resilience when the faction of the MPA is not large. In contrast, when target species has a relatively high migration rate ($\beta_4 = 1.0$), introducing the MPA always increases the relative ecological resilience (Fig. 4b). This qualitative result is kept for the higher migration rates ($\beta_4 \geq 1$). Note that the end points on the curve in Fig. 4 indicate that multiple stable states no longer occur when the fraction of the MPA is over the values of the end points. This suggests that the MPA has an effect of removing multiple stable states.

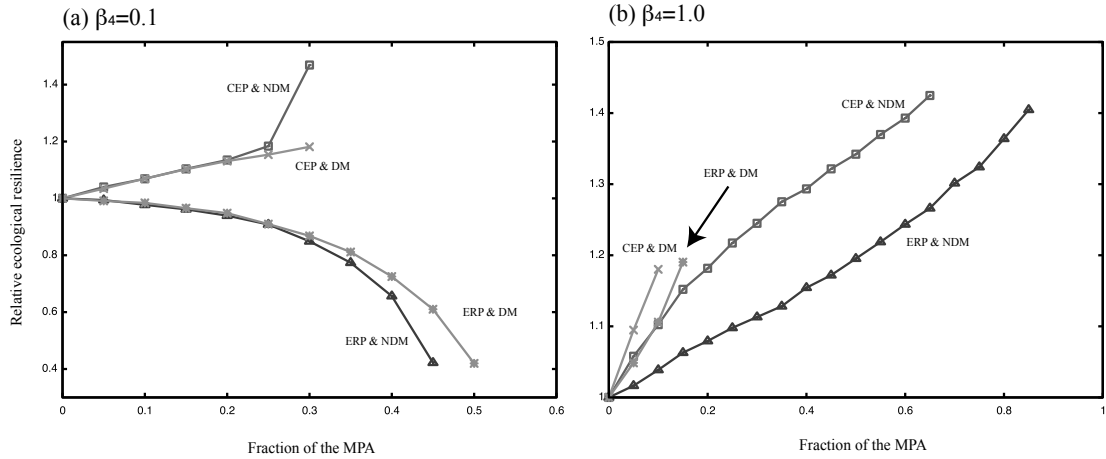

**Figure 4** Ecosystem resilience of upper stable states (>80% value of the uppermost stable states; see text). Each panel represents the effect of the MPA on: (a) species with a very low migration rate ($\beta_4 = 0.1$); (b) species with a low migration rate ($\beta_4 = 1.0$). Strength of density-dependence, *s*, is 2.0 for DM and -0.5 for NDM.

## 4. Discussion

In this paper, we examined the effects of MPAs on ecosystems with multiple stable states. We found that MPAs have two main positive effects for depleted fishing stocks and resilience based ecosystem management. First, the equilibrium population size of a depleted stock tends to recover with the introduction of the MPA. Second, introduction of MPAs can enhance ecological resilience. However, the efficiency of these effects depends heavily on the migration characteristics of the target species, on fishing policies and on the fishing goals. With incorrect management, MPAs might have no effect on the recovery and possibly have a negative effect on the ecological resilience.

The effect of MPA on the recovery of fishing stocks largely depends on the migration ability of the fishing target species. Previous studies have shown that sedentary species receive more conservation benefits from an MPA than migratory species [38-40]. However, we found the opposite result: for a sedentary species, MPAs might be less beneficial for the stock recovery when MPA is small. This suggests that models that do not consider multiple stable states could be misleading for management using MPAs. Nes & Scheffer [41] suggest that spatial heterogeneity lessens the tendency for large-scale catastrophic regime shifts when the species dispersion is low. In the context of the MPA management, which in fact corresponds to introducing spatial heterogeneity into a fishing ground, it may weaken the large-scale drastic recovery, as sedentary species have low contributions to other patches.

The effects of MPAs are also highly dependent on the fishing policies. Our results suggest that when migration rate is very low the transfer fishing effort into fishing grounds due to the creation of MPAs (ERP) is an ineffective or possibly harmful management practice in comparison to no transfer of fishing effort (CEP). The processes of effort redistribution after the institution of MPAs are quite complex and case specific [42-44] but sometimes the converted fishing effort reduces the benefit of the MPA [45]. However, keeping per area fishing efforts constant after introducing MPAs may be difficult because, in most of cases, many fishers are economically or socially dependent on that species. This difficulty might be overcame by, for instance, a fisheries management with feedback control, which redirects fishing effort towards more abundant species in response to changes in populations, rather than reducing the total fishing effort [46].

In the analysis we assume that multiple stable states are introduced by the fishing activity. However, it is also plausible to assume that the ecosystem inherently has multiple stable states when the MPA is absent. In that case, an introduction of the MPA still enforces the ecological resilience although multiple stable states are not disappeared as in Fig. 4 (see Fig. A3 in Appendix). The MPA also shows the similar effect for the recovery of equilibrium population size (see Fig. A4 in Appendix). However, if we apply the initial conditions used in above analysis, the drastic recovery is less likely to occur and the depleted population tends to remain the lower stock level. The initial conditions based on Fig. A4 are set as $\hat{x}_1^0 = R \times 0.3 \hat{x}^U$, $\hat{x}_2^0 = (1-R) \times 0.3 \hat{x}^U$, respectively; these initial values are greater than that used above analysis. This fact indicates that (i) the effect of the MPA could be similar if the population does not totally converge to the lower stable state and (ii) the population recovery may not occur by introducing the MPA unless the population size shows fluctuations if once the population converges to the lower stable state. Note how large initial conditions are needed for the drastic recovery depend on the degree of the ecological resilience.

The manner of fish migration also greatly affects the results of MPA management. Even in well-studied regions, the life-history and movement parameters are not known for more than a handful of species [39]. In addition, for a real marine species, movement manner varies over its life history. A simple formalization of the migration function cannot cover the complexity of every kind of migration pattern [40].

Our model is constructed simply to capture the central processes of fish population dynamics such as density-dependent growth, predation, harvesting and migration. The characteristics of a species are treated as identical over the life history. The fish migrations are also approximated by setting 2-patch environments. Knowledge of migration patterns and parameters of a fishing target species and development of models focusing on the detailed movement patterns would be useful for the establishment of more effective MPAs. For instance, our analysis suggests that the strength of density-dependence $s$ and migration rate $m$ are the key parameters that can change the MPA effects greatly. Therefore, even though the full description of the migration manner is difficult, starting from the estimation of only the key parameters could contribute to more reliable MPA planning.

In this study we do not consider the dynamics of predatory species by assuming the constant predation due to avoiding complexities and uncertainties such as migration manner of predatory species. However, interaction with predatory species is often key to understand population dynamics of concerned species (e.g., [26]) Therefore, including the dynamics of predatory species is needed for future works.

Long-term fluctuations are inherent in marine ecosystems. For example, fluctuations in pelagic species such as Pacific sardine and Atlantic herring have occurred about every 50 to 100 years [47]. In addition, global warming could cause the long-term monotonic environmental changes such as a monotonically increase of sea temperature [48]. These facts suggest the limitation of our model. Because our model do not incorporate the effect of environmental fluctuations, it can only be applied to species that do not undergo environmental fluctuations, species that their biological parameters are not affected even under these perturbations, or the time scale which short enough not to cause parameter changes.

Although our model does not consider environmental fluctuations explicitly, the concept of the ecological resilience offers an index of robustness of the system against such fluctuations because the ecological resilience reflects how large perturbation can absorb without changes in a system function or structure. Because almost species are affected by environmental fluctuation, the ecological resilience does matter. Especially for a species, which is sensitive to perturbations (showing large variance in their biological parameters), small reduction in the ecological resilience has a greater risk of regime shift than other species. Therefore, the ecological resilience is important especially in noisy ecosystems and knowing how large biological parameters fluctuate indicates the relative importance of the ecological resilience for the species.

We revealed that the establishment of the MPA has effects to recover depleted fishing stocks and to enhance ecological resilience, but a reduction of the fishing effort without creating an MPA may also have similar effects (e.g., [12]). However, one of the critical differences between our model and May's model is that in our model the fishing effort can increase after an introduction of the MPA. Therefore the reduction of the ecological resilience can occur in our model. We confirmed numerically that May's model does not show any reduction of the ecological resilience when the fishing effort is decreased.

MPAs can be an effective tool for both recovery of depleted fishing stocks and reinforcement of the ecological resilience. However, if MPAs are managed unsuitably, their creation may have little effect on the recovery of depleted stocks or may even make the existing condition worse. Fisheries managers need to carefully consider use of MPAs, based on the characteristics of the target species and the management goals. If we are interested in a management of depleted fishing stocks, the stock recovery is the primary concern. In such situation, we would focus on the ecosystem after the collapse of a stock. On the one hand, if we are interested in maintenance of ecosystems, ecological resilience may be a more important concern before stock collapse. We may be able to avoid ineffective management through conscientious monitoring of fish migration and by making suitable management rules.

**Acknowledgements**
This work was done with the support of the Grant-in-Aid for Japan Society for the Promotion of Science (JSPS) Fellows and the Grant-in-Aid for Scientific Research (B) to Yoh Iwasa. We thank K. Bessho, Y. Iwasa, K. Noshita, K. Saeki, and Y. Tachiki for their very useful comments.

**Appendix A**

Here, we describe how the density-dependent migratory terms in Eqs. 3 are introduced. We deliberately omit the explanation of other parts on the right-hand side of Eqs. 3 and they are described by a single function $F$ in Appendix A, because they are quite common for ecological model. However, note that dimensions of any parameters used both in the Eqs. 3 and the equations discussed below should not be changed. One of the easy ways to obtain the migratory terms of Eqs. 3 is to transform from the population dynamics in number into that of in density. When the population is measured in number, the population dynamics in the patch 1 (fishing ground) is represented as follows:

$$\frac{dx_1}{dt} = F(x_1, R) + m\left[\left(\frac{x_2}{RK}\right)^s (1-R)x_2 - \left(\frac{x_1}{(1-R)K}\right)^s R x_1\right], \quad (A.1)$$

where, $x_1$ and $x_2$ are the number of individuals in the patch 1 and 2 (MPA),

respectively. $F(x_1, R)$ is the rate of change of $x_1$ apart from emigration and immigration effects. $m$ is the migration rate, $K$ is the carrying capacity measured in unit of number per area (same as Eq. 1), $s$ is the strength of density-dependence of migration [1], $R$ is the fraction of the MPA and $1-R$ is the fraction of the fishing ground. The first and second terms in the square brackets represent the immigration from the patch 2 and the emigration to the patch 2, respectively. When the population "density" affect the rate of migration of an individual, both immigration and emigration rates can be represented by the multiplication of the inherent migration rate, the density effect in the current location, the fraction of the outside patch and the number of individuals in the current patch. To transform the variable of population number $x_1$ into the population density $X_1$, where $X_1 = x_1 / (\text{area of patch})$, in Eq (A.1), we divide the both sides of Eq. (A.1) by the area of patch 1, $A(1-R)$, where $A$ is the area of the concerned space and we assume it is unity). Then we obtain the migratory term of Eq. (3a). In the same manner, we can also obtain the migratory term of Eq. (3b).

Next, we introduce the non-dimensional form of Eqs. (3). By rescaling parameters $\hat{x}_i = X_i / K$, $\tau = rt$, $\beta_1 = b/K$, $\beta_2 = aP/(rK)$, $\beta_3 = qE/(rK)$, and $\beta_4 = m/r$, we obtain Eqs. (4). To emphasize the fraction of the MPA, $R$, we left $R$ in the harvest and migratory terms.

**Reference**

[1] Amarasekare, P., The role of density-dependent dispersal in source-sink dynamics, J. Theor. Biol. 226 (2004) 159-168.

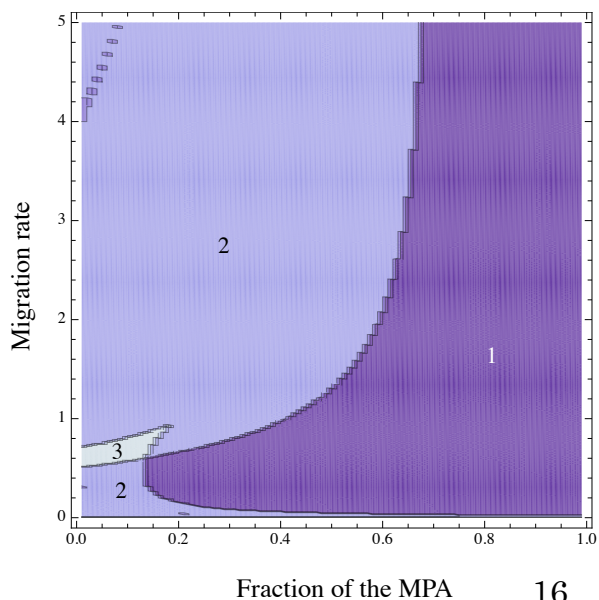

**Figure A1** The number of stable equilibrium in the parameter space. We assume CEP ($\sigma = 0$) and DM ($s = 1$).

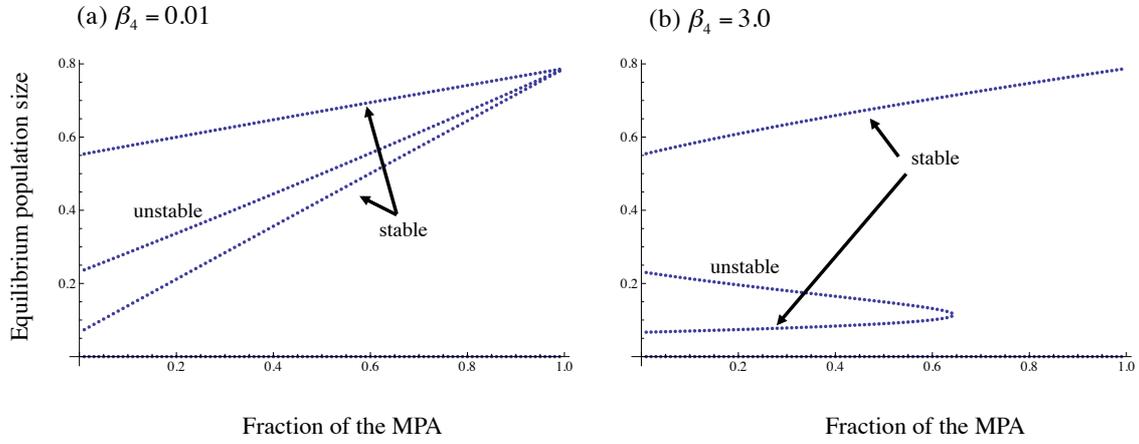

**Figure A2** The plausible equilibrium population size. We assume CEP ($\sigma = 0$) and DM ($s = 1$). The migration rate $\beta_4$ is (a) 0.01 and (b) 3.0.

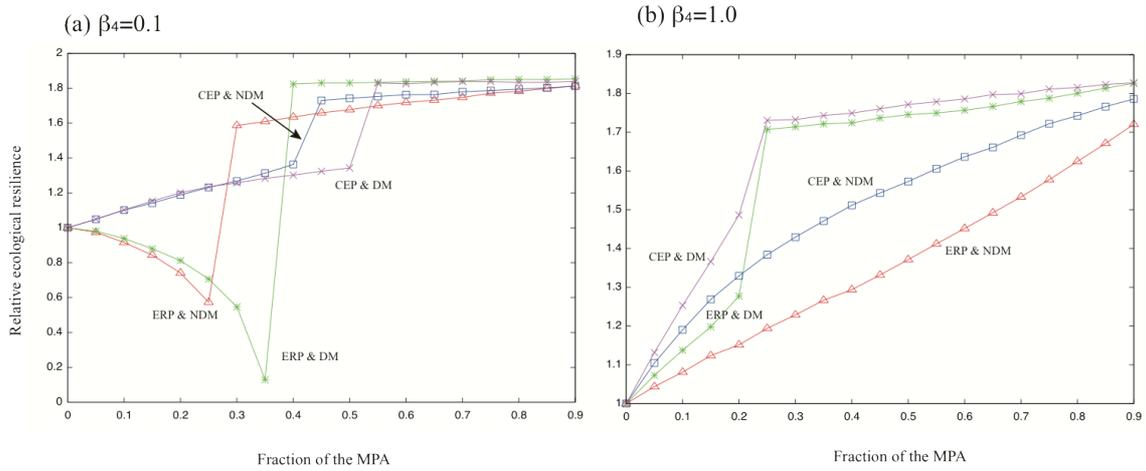

**Figure A3** Ecosystem resilience of upper stable states (>80% value of the uppermost stable states) in an ecosystems where inherently multiple stable states exist. Each panel represents the effect of the MPA on: (a) species with a very low migration rate ($\beta_4 = 0.1$); (b) species with a low migration rate ($\beta_4 = 1.0$). Strength of density-dependence, $s$, is 2.0 for DM and -0.5 for NDM.

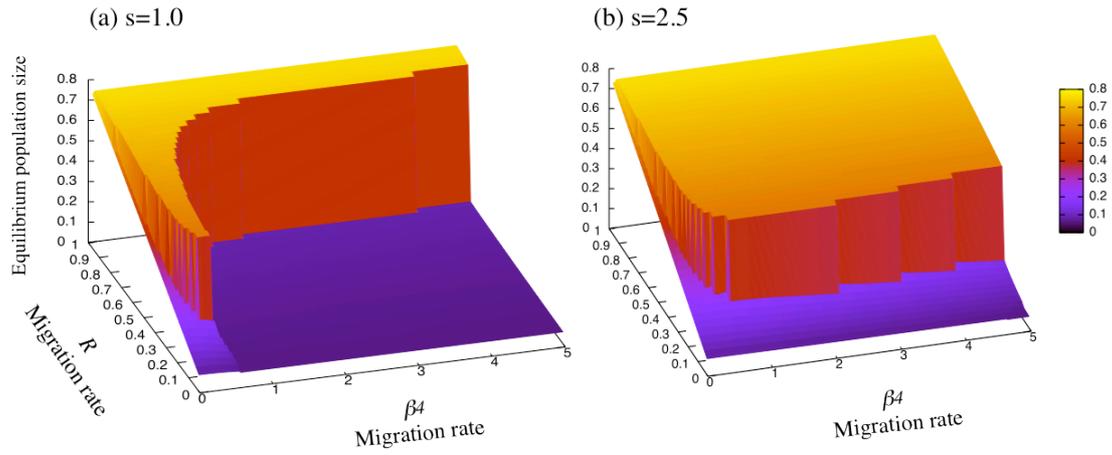

**Figure A4** Effects of introducing the MPA on equilibrium population size in an ecosystems where inherently multiple stable states exist. Colors represent equilibrium population sizes. We assume CEP ($\sigma = 0$) and DM ($s > 0$). The strength of density-dependent migration differs in (a) and (b). The recovery of equilibrium population size due to the MPA depends on whether the dynamics starting from initial population size converge to upper stable equilibrium or not.